\RequirePackage{lineno}
\documentclass[aps,prd,twocolumn,showpacs,superscriptaddress,groupedaddress]{revtex4}  
\usepackage{graphicx}  
\usepackage{dcolumn}   
\usepackage{bm}        
\usepackage{amssymb}   
\usepackage{amsmath}
\usepackage{array}
\newcommand{\PreserveBackslash}[1]{\let\temp=\\#1\let\\=\temp}
\newcolumntype{C}[1]{>{\PreserveBackslash\centering}p{#1}}
\newcolumntype{R}[1]{>{\PreserveBackslash\raggedleft}p{#1}}
\newcolumntype{L}[1]{>{\PreserveBackslash\raggedright}p{#1}}

\begin{document}
\title{First measurement of radioactive isotope production through cosmic-ray muon spallation in Super-Kamiokande IV}
\newcommand{\AFFicrr}{\affiliation{Kamioka Observatory, Institute for Cosmic Ray Research, University of Tokyo, Kamioka, Gifu 506-1205, Japan}}
\newcommand{\AFFkashiwa}{\affiliation{Research Center for Cosmic Neutrinos, Institute for Cosmic Ray Research, University of Tokyo, Kashiwa, Chiba 277-8582, Japan}}
\newcommand{\AFFipmu}{\affiliation{Kavli Institute for the Physics and
Mathematics of the Universe (WPI), The University of Tokyo Institutes for Advanced Study,
University of Tokyo, Kashiwa, Chiba 277-8582, Japan }}
\newcommand{\AFFmad}{\affiliation{Department of Theoretical Physics, University Autonoma Madrid, 28049 Madrid, Spain}}
\newcommand{\AFFubc}{\affiliation{Department of Physics and Astronomy, University of British Columbia, Vancouver, BC, V6T1Z4, Canada}}
\newcommand{\AFFbu}{\affiliation{Department of Physics, Boston University, Boston, MA 02215, USA}}
\newcommand{\AFFbnl}{\affiliation{Physics Department, Brookhaven National Laboratory, Upton, NY 11973, USA}}
\newcommand{\AFFuci}{\affiliation{Department of Physics and Astronomy, University of California, Irvine, Irvine, CA 92697-4575, USA }}
\newcommand{\AFFcsu}{\affiliation{Department of Physics, California State University, Dominguez Hills, Carson, CA 90747, USA}}
\newcommand{\AFFcnm}{\affiliation{Department of Physics, Chonnam National University, Kwangju 500-757, Korea}}
\newcommand{\AFFduke}{\affiliation{Department of Physics, Duke University, Durham NC 27708, USA}}
\newcommand{\AFFfukuoka}{\affiliation{Junior College, Fukuoka Institute of Technology, Fukuoka, Fukuoka 811-0295, Japan}}
\newcommand{\AFFgifu}{\affiliation{Department of Physics, Gifu University, Gifu, Gifu 501-1193, Japan}}
\newcommand{\AFFgist}{\affiliation{GIST College, Gwangju Institute of Science and Technology, Gwangju 500-712, Korea}}
\newcommand{\AFFuh}{\affiliation{Department of Physics and Astronomy, University of Hawaii, Honolulu, HI 96822, USA}}
\newcommand{\AFFkek}{\affiliation{High Energy Accelerator Research Organization (KEK), Tsukuba, Ibaraki 305-0801, Japan }}
\newcommand{\AFFkobe}{\affiliation{Department of Physics, Kobe University, Kobe, Hyogo 657-8501, Japan}}
\newcommand{\AFFkyoto}{\affiliation{Department of Physics, Kyoto University, Kyoto, Kyoto 606-8502, Japan}}
\newcommand{\AFFmiyagi}{\affiliation{Department of Physics, Miyagi University of Education, Sendai, Miyagi 980-0845, Japan}}
\newcommand{\AFFnagoya}{\affiliation{Solar Terrestrial Environment Laboratory, Nagoya University, Nagoya, Aichi 464-8602, Japan}}
\newcommand{\AFFpol}{\affiliation{National Centre For Nuclear Research, 00-681 Warsaw, Poland}}
\newcommand{\AFFsuny}{\affiliation{Department of Physics and Astronomy, State University of New York at Stony Brook, NY 11794-3800, USA}}
\newcommand{\AFFokayama}{\affiliation{Department of Physics, Okayama University, Okayama, Okayama 700-8530, Japan }}
\newcommand{\AFFosaka}{\affiliation{Department of Physics, Osaka University, Toyonaka, Osaka 560-0043, Japan}}
\newcommand{\AFFregina}{\affiliation{Department of Physics, University of Regina, 3737 Wascana Parkway, Regina, SK, S4SOA2, Canada}}
\newcommand{\AFFseoul}{\affiliation{Department of Physics, Seoul National University, Seoul 151-742, Korea}}
\newcommand{\AFFshizuokasc}{\affiliation{Department of Informatics in
Social Welfare, Shizuoka University of Welfare, Yaizu, Shizuoka, 425-8611, Japan}}
\newcommand{\AFFskk}{\affiliation{Department of Physics, Sungkyunkwan University, Suwon 440-746, Korea}}
\newcommand{\AFFtokyo}{\affiliation{The University of Tokyo, Bunkyo, Tokyo 113-0033, Japan }}
\newcommand{\AFFtoronto}{\affiliation{Department of Physics, University of Toronto, 60 St., Toronto, Ontario, M5S1A7, Canada }}
\newcommand{\AFFtriumf}{\affiliation{TRIUMF, 4004 Wesbrook Mall, Vancouver, BC, V6T2A3, Canada }}
\newcommand{\AFFtokai}{\affiliation{Department of Physics, Tokai University, Hiratsuka, Kanagawa 259-1292, Japan}}
\newcommand{\AFFtsinghua}{\affiliation{Department of Engineering Physics, Tsinghua University, Beijing, 100084, China}}
\newcommand{\AFFuw}{\affiliation{Department of Physics, University of Washington, Seattle, WA 98195-1560, USA}}

\AFFicrr
\AFFkashiwa
\AFFmad
\AFFbu
\AFFubc
\AFFbnl
\AFFuci
\AFFcsu
\AFFcnm
\AFFduke
\AFFfukuoka
\AFFgifu
\AFFgist
\AFFuh
\AFFkek
\AFFkobe
\AFFkyoto
\AFFmiyagi
\AFFnagoya
\AFFpol
\AFFsuny
\AFFokayama
\AFFosaka
\AFFregina
\AFFseoul
\AFFshizuokasc
\AFFskk
\AFFtokai
\AFFtokyo
\AFFipmu
\AFFtoronto
\AFFtriumf
\AFFtsinghua
\AFFuw

\author{Y.~Zhang}
\AFFtsinghua
\author{K.~Abe}
\AFFicrr
\AFFipmu
\author{Y.~Haga}
\AFFicrr
\author{Y.~Hayato}
\AFFicrr
\AFFipmu
\author{M.~Ikeda}
\AFFicrr
\author{K.~Iyogi}
\AFFicrr
\author{J.~Kameda}
\author{Y.~Kishimoto}
\author{M.~Miura}
\author{S.~Moriyama}
\author{M.~Nakahata}
\AFFicrr
\AFFipmu
% ...

\collaboration{The Super-Kamiokande Collaboration}
\noaffiliation
\date{\today}

\begin{abstract}
Cosmic-ray-muon spallation-induced radioactive isotopes with $\beta$ decays are one of the major backgrounds for solar, reactor, and supernova relic neutrino experiments.  Unlike in scintillator, production yields for cosmogenic backgrounds in water have not been exclusively measured before, yet they are becoming more and more important in next generation neutrino experiments designed to search for rare signals. We have analyzed the low-energy trigger data collected at Super-Kamiokande-IV in order to determine the production rates of $^{12}$B, $^{12}$N, $^{16}$N, $^{11}$Be, $^9$Li, $^8$He, $^9$C, $^8$Li, $^8$B and $^{15}$C. These rates were extracted from fits to time differences between parent muons and subsequent daughter $\beta$'s by fixing the known isotope lifetimes. Since $^9$Li can fake an inverse-beta-decay reaction chain via a $\beta + n$ cascade decay, producing an irreducible background with detected energy up to a dozen MeV, a dedicated study is needed for evaluating its impact on future measurements; the application of a neutron tagging technique using correlated triggers was found to improve this $^9$Li  measurement. The measured yields were generally found to be comparable with theoretical calculations, except the cases of the isotopes $^8$Li/$^8$B and $^9$Li.
\end{abstract}
\maketitle

\section{Introduction}
Cosmic-ray muon spallation induced radioactive isotopes with $\beta$ decays are one of the major backgrounds for solar, reactor, and supernova relic neutrino experiments. Although the production yields for spallation backgrounds have been extensively measured in liquid scintillator experiments~\cite{hagner2000, kamland2010, bxno2013}, there has been a lack of measurements of these isotopes produced in water. In this case, it is understood that cosmic-ray muons and their subsequent showers 
interact with $^{16}$O nuclei to produce isotopes, which usually have a lifetime up to tens of seconds and can undergo a cascade decay with $\beta$ energy ranging from a few MeV to a dozen MeV. For neutrino interactions in this energy range, these events can form a background in planned next-generation water Cherenkov detectors~\cite{gdzooks!, hk, watchman, memphys, lbne}, which are designed for detecting signals such as solar neutrino-electron elastic scattering or inverse beta decay (IBD; $\bar{\nu}_e + p \rightarrow e^+ + n$) of anti-neutrinos from reactors or core-collapse supernovae. 

Previous studies at Super-Kamiokande (SK) -- where the cosmic-ray muon rate is around 2 Hz --  developed cuts to remove the spallation backgrounds produced in water by using the event time offset and distance to the preceding muon as well as the muon's light production in relation to its path length~\cite{Fukuda2001, Hosaka2006, Abe2011, Bays2012}. However, many long-lived isotopes still survived and affected the search for a low-energy upturn of the solar $^8$B neutrino spectrum as well as the hunt for evidence of supernova relic neutrinos (SRNs),  also known as diffuse supernova neutrino background (DSNB), emitted from past core-collapse supernovae. Recently, theoretical progress has been made in the calculation of the production and properties of these backgrounds in water, shedding light on practical improvements in spallation background rejection \cite{li2014, li2015, li20152}. Providing experimental data should help improve knowledge in this area.

In addition, certain spallation products in water are of particular interest for anti-neutrino detection, as decays from these isotopes can mimic the IBD signature.  This affects the search for reactor neutrinos as well as for SRNs. The SRN search is particularly vulnerable to the spallation background since its expected signal event rate is quite low. We find that of the potential spallation isotopes only the decay of $^9$Li can have a significant contribution through the $\beta$ energy tail to the \lq \lq golden\rq \rq ~SRN search window of 9.5$-$29.5 MeV in positron kinetic energy, in which the contribution from both reactor neutrino and atmospheric neutrino backgrounds is relatively small. Theoretically, the $^9$Li isotope is known to be not directly produced by the primary muon, but rather by the secondary $\pi^-$ interacting with the $^{16}$O  nucleus: $\pi^- + ^{16}\mbox{O} \rightarrow \alpha + 2p + n + ^{9}\mbox{Li}$, where  $^9$Li has a lifetime of 0.26 s and decays into $\beta+n$ with a probability of ($50.8\pm0.9$)\%~\cite{Tilley}. This feature can be exploited to separate the $^9$Li candidates from other isotope events and give a better measurement on the corresponding yield, if the delayed neutrons are tagged as was done in the SK-IV SRN search~\cite{Zhang2015}. 
 
In this article, the study of cosmogenic isotopes produced in SK-IV is described in detail.
Section~\ref{sec:SK} introduces the SK detector and performance. 
Section~\ref{sec:muons} describes the definition and classification of cosmic-ray muons. 
Section~\ref{sec:yielddt}  presents the measurement of production rates of spallation isotopes from fits to the lifetime distributions. Section~\ref{sec:yieldntag} describes a modification of the neutron tagging and gives the improved measurement for $^9$Li. Section~\ref{sec:discuss} gives results and discussions, focusing on a comparison with the theoretical calculations and the impact on future experiments using the water Cherenkov technique. Section~\ref{sec:summary} contains a summary. 

\section{EXPERIMENT}
\label{sec:SK}
The SK experiment, a cylindrical, 50-kton water Cherenkov detector with 2,700 meter water equivalent overburden, consists of two volumes: an outer detector (OD) and an inner detector (ID)~\cite{SK}. 
The OD is viewed by 1,885 outward-pointing
photomultiplier tubes (PMTs), which are used to tag charged particles 
originating from outside (or exiting) the ID. An OD trigger is issued when the number of OD PMT hits within 200 ns exceeds 22.  The ID 
has a diameter of 33.8 m and a height of 36.2 m and is
surrounded by 11,129 inward-pointing PMTs, which are used to detect the Cherenkov light
from charged particles traveling in the ID.  

At SK, low-energy events are triggered by the number of coincident ID PMT hits within 200 ns, in which 
an ID PMT hit is defined as one having a recorded charge over 0.25 photo-electrons. 
In the summer of 2008, a new electronics and online system was successfully installed to enhance  the data-processing capability of SK~\cite{SK4ONLINE}.  
The running period after that time is referred to as SK-IV, during which the trigger threshold for 
low-energy events was consequently lowered to 3.5 MeV.
Physics analyses for solar neutrinos and other low energy phenomena are performed with this trigger. 
In addition, the new online system allowed the introduction of a correlated trigger scheme.   
Low-energy events with a threshold over 9.5 MeV (reduced to 7.5 MeV after the summer of 2011) were 
followed by a forced trigger 500 $\mu$s long to record all the PMT hits, including those fired by the subtle 2.2 MeV $\gamma$ emitted 
from a neutron capture on hydrogen ($n + p \rightarrow d + \gamma$). 
Due to the readout configuration of the new online system, data for normal triggers were taken for 40 $\mu$s ranging 
from -5 to 35 $\mu$s with respect to the time when the triggers were issued. 
Therefore, delayed 2.2 MeV $\gamma$ signals can be searched for up to 
535 $\mu$s after the primary event time.   
This study used data collected from October 2008 to October 2014. 

Reconstruction, energy, and position calibrations for low-energy events in SK are carefully performed 
as described in Ref.~\cite{Abe2011}. 
The vertex resolution is about 65 cm for a 7.5 MeV electron or positron (indistinguishable in SK) and improves as energy increases. The energy of an event is reconstructed
from the number of detected Cherenkov photons. That number (approximately proportional to the
energy of a charged particle) is estimated by the number of PMT hits within 50 ns and corrected for
late hits, dark noise, bad PMTs, photocathode coverage and water transparency.
In this paper, the energy we use is the kinetic energy of the electron/positron.

\section{COSMIC RAY MUON DATA}
\label{sec:muons}
\begin{figure}[!t]
   \centering
  \includegraphics[width=0.9\columnwidth]{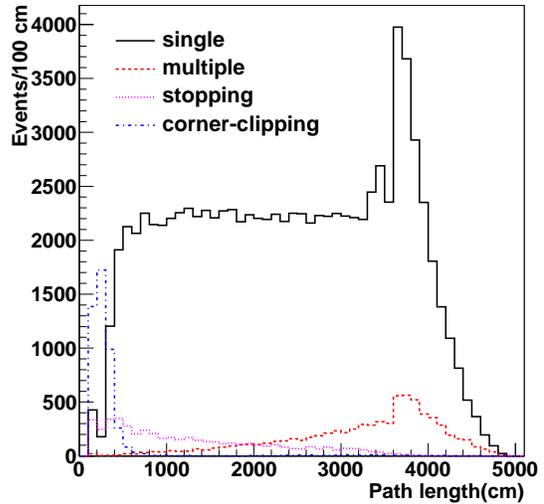}
   \renewcommand{\figurename}{Fig.}
   \caption{The path-length distributions for single through-going (solid), multiple (dashed), stopping (dotted) and corner-clipping (dot-dashed) muons, respectively. }
   \label{fig:pl}
 \end{figure}

The reconstruction of cosmic-ray muons is of critical importance in both tagging spallation backgrounds
and selecting cosmic $\mu$-e decays. Muon candidates are required
to have both an ID and an OD trigger. In addition, the number of photo-electrons (p.e.s) collected
should be greater than 1,000 ($\sim$140 MeV) in the ID. In this study, we only analyzed muons within $\pm$30 seconds around the low-energy events. These muon candidates were reconstructed with a dedicated muon fitter. 
Details about the algorithm of the muon fitter can be found elsewhere~\cite{muon1, muon2}.
The muon fitter categorized the cosmic ray muons into four classes 
\begin{itemize} 
\item [(I)] Single through-going muons: those with a track penetrating the ID. (84.1$\pm$0.3)\% of all muons
\item [(II)] Multiple muons: those with several parallel muon tracks in the ID. (6.9$\pm$0.2)\%
\item [(III)] Stopping muons: those with a track that entered but did not leave the ID. (4.6$\pm$0.2)\%
\item [(IV)] Corner-clipping muons: those with a track length shorter than 7 m inside the ID. (4.1$\pm$0.1)\%
\end{itemize}
The remaining 0.3\% of the muon events were not identified by the muon fitting algorithm 
due to an insufficient number of PMT hits after the selection cuts. These events were referred to as non-fitted muons. 
In addition, we observed that 1.2\% of the muons among the four classes had a poor quality of fit and were thus classified as poorly-fitted muons. The non-fitted and poorly-fitted muons were still used in the studies of the 
time distributions even though the transverse distance cut defined in the following section could not 
be applied, since the entrance position for cosmic-ray muon was not determined by the muon fitter.     
Here we only used the most probable track information for multiple muons in the analysis.
Fig.~\ref{fig:pl} shows the path-length distribution for each muon category. The distributions reflect the effects due to the geometry of the detector and the zenith angle distribution of the cosmic-ray muons. The peaks around 36 m are from muons traveling vertically downward through the detector. The muon rate $R_{\mu}$ and the average path length for muons $L_{\mu}$ were found to be $R_{\mu} = 2.00 \pm 0.01$ Hz and $L_{\mu} = 2,200 \pm 4 $ cm, respectively.

\section{MEASUREMENT OF ISOTOPE RATES}
\label{sec:yielddt}
The candidates for radioactive decays from spallation events were selected using cuts on both the temporal and spatial relationship
between the low-energy events and the preceding muons. The live time of the data set used for this study is 1,890 days. The time distribution of low-energy events up to 30 s after each muon was used to extract the production rate of radioactive nuclei (in the spallation backgrounds).
Table~\ref{tab:isotope} lists the possible radioactive isotopes induced by cosmic-ray muon spallation at SK~\cite{nndc, li2014, duke}. \begin{table*}[thbp]
  \caption[Radioactive isotopes induced by cosmic ray muons at SK]
          {Possible radioactive isotopes induced by cosmic-ray muon spallation at SK~\cite{nndc, li2014, duke}. The fourth
          column lists the end point kinetic energy ($E_{\text{kin.}}$). The fifth
          column lists the primary generation process of the radioactive isotopes.}
  \label{tab:isotope}
  \centering
  \begin{tabular}{lcccr} \hline
    \hline Radioactive isotope & $\tau$ (s) & Decay mode &  $E_{\text{kin.}}$ (MeV)  & Primary process\\
    \hline ${}^{11}$Be & 19.9   &  $\beta^-$   & 11.51   &  $^{16}\text{O}(n, \alpha+2p){}^{11}$Be\\
                          &        &  $\beta^- \gamma$   & 9.41+2.1($\gamma$) & \\
           ${}^{16}$N  & 10.3  &  $\beta^-$   & 10.44 & $^{16}\text{O}(n,p){}^{16}$N \\
                          &        &  $\beta^- \gamma$   & 4.27+6.13($\gamma$) & \\
           ${}^{15}$C  & 3.53  &  $\beta^-$    & 9.77 & $^{16}\text{O}(n,2p){}^{15}$C\\
                          &        &   $\beta^- \gamma$  & 4.51+5.30($\gamma$) & \\
           ${}^{8}$Li  & 1.21   &  $\beta^-$   & $\sim$13.0 & $^{16}\text{O}(\pi^-, \alpha+{}^2\text{H}+p+n){}^{8}$Li\\
           ${}^{8}$B   & 1.11   &  $\beta^+$   & $\sim$13.9 & $^{16}\text{O}(\pi^+, \alpha+2p+2n){}^{8}$B \\
           ${}^{16}$C  & 1.08  &  $\beta^-+n$ & $\sim$4 & $^{18}\text{O}(\pi^-,n+p){}^{16}$C\\
           ${}^{9}$Li  & 0.26 &  $\beta^-$   & 13.6 & $^{16}\text{O}(\pi^-, \alpha+2p+n){}^{9}$Li \\
                          &        &  $\beta^-+n$ & $\sim$10 & \\
           ${}^{9}$C   & 0.18 &  $\beta^++p$ & 3$\sim$15 & $^{16}\text{O}(n, \alpha+4n){}^{9}$C\\
           ${}^{8}$He  & 0.17 &  $\beta^- \gamma$   & 9.67+0.98($\gamma$) & $^{16}\text{O}(\pi^-, {}^3\text{H}+4p+n){}^{8}$He\\
                          &        &  $\beta^-+n$ &  & \\
           ${}^{12}$Be & 0.034 &  $\beta^-$   & 11.71 & $^{18}\text{O}(\pi^-, \alpha+p+n){}^{12}$Be\\
           ${}^{12}$B  & 0.029 &  $\beta^-$   & 13.37 & $^{16}\text{O}(n, \alpha+p){}^{12}$B \\
           ${}^{13}$B  & 0.025 &  $\beta^-$   & 13.44 & $^{16}\text{O}(\pi^-, 2p+n){}^{13}$B\\
           ${}^{14}$B  & 0.02 &  $\beta^- \gamma$   & 14.55+6.09($\gamma$) & $^{16}\text{O}(n, 3p){}^{14}$B\\
           ${}^{12}$N  & 0.016 &  $\beta^+$   & 16.38 & $^{16}\text{O}(\pi^+, 2p+2n){}^{12}$N\\
            ${}^{13}$O  & 0.013 &  $\beta^++p$ & 8$\sim$14 & $^{16}\text{O}(\mu^-, \mu^- + p + 2n + \pi^-){}^{13}$O\\
            ${}^{11}$Li & 0.012 &  $\beta^-$   & 20.62  & $^{16}\text{O}(\pi^+, 5p+\pi^0+\pi^+){}^{11}$Li\\
                          &        &  $\beta^-+n$ & $\sim$16 & \\
    \hline \hline
  \end{tabular}
\end{table*}

\subsection{Data reduction}
The cuts for selecting the signal of spallation products from the low-energy events are described in this section.
\subsubsection{First reduction}
We applied a first event reduction to remove non-physical events using the following criteria \cite{Hosaka2006, Bays2012}. 
\begin{itemize}
\item [(I)] Events with a noise ratio greater than 0.55 were removed, where the noise ratio is defined as the fraction of PMT hits with charge lower than 0.5 photo-electrons compared to the total number of PMT hits.
\item [(II)] Events with a reconstructed position less than 2 m from the ID wall were removed.
\item [(III)] Events within 50 $\mu$s of a preceding muon were rejected in order to remove cosmic $\mu$-e decays, as well as to eliminate events due to after-pulsing of the PMTs.
\end{itemize}
Using a Geant 3.21-based Monte Carlo (MC) simulation package \cite{Geant3}, the efficiency for the dominant spallation products passing the first reduction cuts was estimated to be greater than (99$\pm$1)\%.

\subsubsection{Energy}
For this analysis, events were required to have a detected energy above 6 MeV since below this energy threshold non-spallation events dominate. The efficiency for spallation events passing the energy cut is assessed individually for each isotope using the corresponding $\beta$ energy spectra. In this article, all the theoretical spectra for the isotopes were obtained from Ref.~\cite{duke}. It should be noted that the detected energy might include energy from $\gamma$'s in addition to the expected $\beta$ energy. This was taken into account in the simulation.

\subsubsection{Distance and Time}
\label{subsec:disdt}
In previous SK analyses, the transverse distance (\emph{lt}) between a muon track and a low-energy event was used as a powerful cut to remove the spallation background. 
This cut was re-investigated for spallation events.  In this analysis, 
a correlated muon event was defined as a muon recorded less than 30 s prior to a low-energy event, while an uncorrelated muon event was defined as a muon recorded less than 30 s after a low-energy event. The muon along with the low-energy events corresponding to the correlated and uncorrelated muon classes comprise the spallation and non-spallation samples, respectively.
The absolute time difference \emph{dt} was calculated from the muon time and 
the low-energy-event time. Under these definitions, we statistically obtained the distributions for the 
spallation signal events from a subtraction of the non-spallation sample from that of the spallation sample.  
This helped us to have a better understanding to the mechanism of muon spallation. 

\begin{figure*}[!t]
   \centering
  \includegraphics[width=2.0\columnwidth]{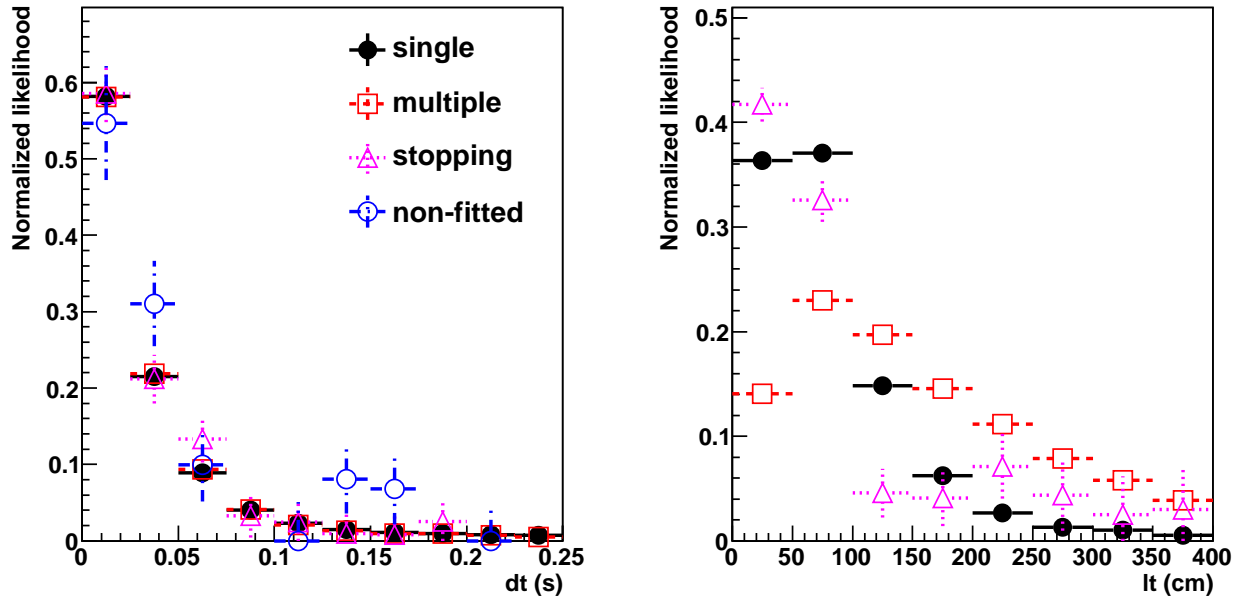}
   \renewcommand{\figurename}{Fig.}
   \caption{The normalized \emph{dt} (left panel) and \emph{lt} (right panel) 
   distributions for the spallation signal events with single through-going muons (solid points), multiple muons 
   (open squares), stopping muons (open triangles) and non-fitted muons (open circles), respectively. 
   The $dt$ and $lt$ distributions for corner-clipping muons are flat and 
   are therefore not shown. Since there is no muon 
   tracking information for non-fitted muons, no corresponding $lt$ distribution is shown. 
   See text for more details.}
   \label{fig:dtdl}
 \end{figure*}

Fig.~\ref{fig:dtdl} shows the subtracted distributions of $dt$ and $lt$ for all the muon categories 
except for the corner clipping muons. We normalized the distributions to unity in order to 
give a shape comparison. Since the isotopes produced by corner-clipping muons either were removed by data reduction or escaped from the ID, the $dt$ and $lt$ distributions for these muons were dominated by the non-spallation muons.    
The $dt$ distributions are similar for each fitted muon class, and so we conclude that there is no correlation between the type of muon and the resulting spallation isotope. However, we observed that the $lt$ distribution for multiple muons was broader than the others. This was understood to be due to the choice of evaluating $lt$ from the most probable track
out of several parallel muon tracks in each bundle.  Consequently, a cut on $lt$ had different impacts on the muons and should be studied carefully. 

To reject non-spallation muons, we required that $lt<$ 200 cm for all the muons
except for the non-fitted and poorly-fitted muons, which had no or no reliable muon track information as described in Section~\ref{sec:muons}.
The efficiency of the $lt$ cut for all the combined spallation muon categories was directly evaluated from the data.  The total number of spallation signal events was obtained from 
the number of events in the spallation sample subtracted by that in the non-spallation sample. The fraction of spallation signal events surviving  
the $lt$ cut gave the efficiency to be ($78.8\pm0.6$)\%, where the error is statistical only. 
We investigated the systematics by applying different $dt$ cuts and checked if these cuts could affect the contribution from the isotopes with different lifetimes. We found 
the relative variation was within 4.8\%, which was assigned as the systematic uncertainty for the $lt$ cut.     
\subsection{Production rates}
We used the spallation sample 
to measure the production rates of the isotopes. 
Table~\ref{tab:cmpspgl} lists the measured major radioactive isotopes induced by the 
spallation at SK with lifetimes
varying from 0.016 s to 19.9 s (see Table~\ref{tab:isotope}).
Fitting the time distributions using discrete time ranges simplified the procedure for extracting the initial production rates of the spallation isotopes.

\subsubsection{Fitting formula}
Assuming the initial amount of $i$-th radioactive isotope as $N_i$, the $dt$ spectrum can be described with the following function: 
\begin{equation}
\label{eq:fitfunction}
F(t)=\sum_i \frac{N_i}{\tau_i}e^{-dt/\tau_i} + const,
\end{equation}
where $\tau_i$ is the lifetime of the $i$-th radioactive isotope and \emph{const} is
the expected number of accidental events, which give a flat time distribution. We can obtain each $N_i$ value by fitting the experimental $dt$ spectrum with the function.

\subsubsection{Fits to sub-time ranges}
Since the lifetimes of the isotopes differ significantly, we can divide the full time range into four parts and obtain $N_i$'s which can then be used as initial values for the final fit to the full time range.
The sub-time ranges given below are arranged according to the fitting sequence.  
\begin{itemize}
\item [(I)] Time range from 50 $\mu$s to 0.1 s:  $i={}^{12}$B and ${}^{12}$N.
\item [(II)] Time range (6$-$30) s:  $i={}^{16}$N and ${}^{11}$Be. 
\item [(III)] Time range (0.1$-$0.8) s: $i={}^{9}$Li,  ${}^{8}$He/${}^{9}$C and ${}^{8}$Li/${}^{8}$B by fixing the $N_i$'s of ${}^{12}$B, ${}^{12}$N, ${}^{16}$N and ${}^{11}$Be obtained above. Since the $N_i$ of ${}^{8}$He could not be decoupled from that of ${}^{9}$C due to their similar lifetimes as shown in Table~\ref{tab:isotope}, only the sum of both with an equal fraction was fitted. 
The same treatment was also applied to ${}^{8}$Li and ${}^{8}$B.   
\item [(IV)] Time range (0.8$-$6) s: $i=N_i$'s of ${}^{15}$C and ${}^{16}$N were obtained by fixing
the $N_i$ of ${}^{8}$Li/${}^{8}$B obtained from the measurement in the range of 0.1$-$0.8 s. 
\end{itemize}

\subsubsection{Fit to full time range}
\begin{figure*}[!t]
   \centering
  \includegraphics[width=2.0\columnwidth]{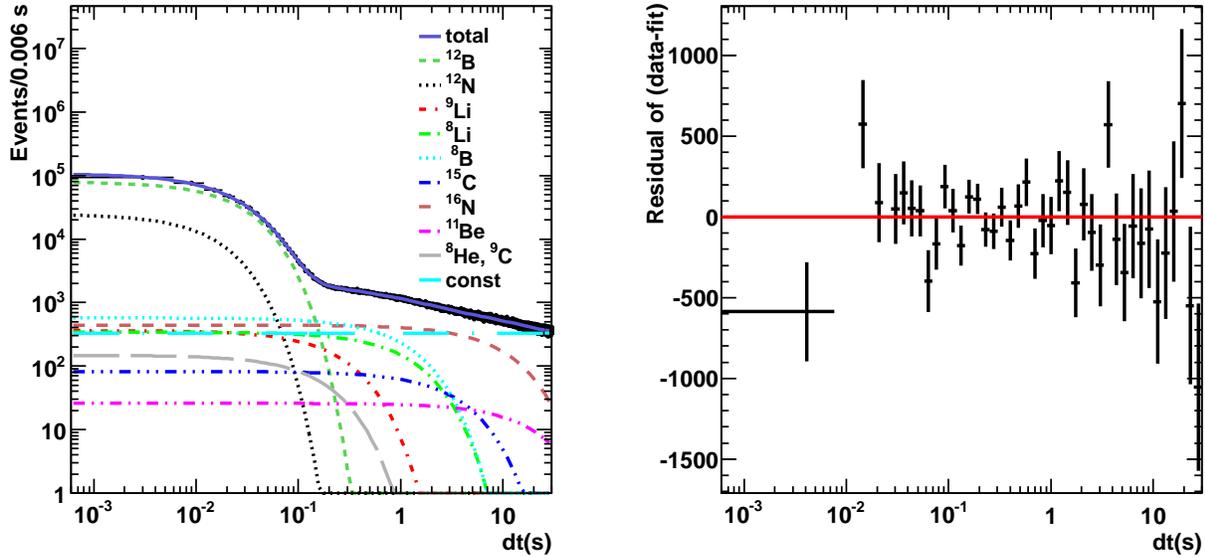}
  \renewcommand{\figurename}{Fig.}
   \caption{Left panel: The fit result in the full time range 
   with a $\chi^2$/ndf=5,167.7/4,991 using the low energy data above 6 MeV.
The points represent the data. The lines denote the radioactive isotope contributions as obtained from the fit. The flat line represents the constant rate of the accidental events. Right panel: The fit residual (data$-$fit) as a
function of time difference ($dt$). The result is gaussian distributed around zero. For the purpose of display, bins are
combined for equal-bin-width in log scale.}
   \label{fig:global}
 \end{figure*}

After all the initial $N_i$'s of radioactive isotopes were obtained from the separated time distributions, they were used as inputs to a fit to the full time range,
where the $N_i$'s of the isotopes were completely free. 
Fig.~\ref{fig:global} shows the fit result with parameters for the nine isotopes and a parameter for the accidental background. The $\chi^2$/ndf was 5,167.7/4,991, corresponding to a $p$$-$value of 4.0\%. To
further illustrate the fit quality, we also show the fit
residual (data $-$ fit) in Fig.~\ref{fig:global}. The result is gaussian distributed around zero. The small $p-$value was due to the fact that some points have large statistical deviation as shown in Fig.~\ref{fig:global} (right panel).

\subsubsection{Production rates}
\label{subsubsec:produc}
With the fitted $N_i$ for each isotope, 
the corresponding production rate $R_i$ was calculated by
\begin{equation}
\label{eq:RI}
R_i= \frac{N_i}{{FV} \cdot {T}\cdot\epsilon_i},
\end{equation}
where $FV$ is the effective fiducial mass, which was 22.5 kton. $\epsilon_i$ is the 
selection efficiency, including the contribution from the first reduction, energy cut and $lt$ cut, for the $i$-th radioactive isotope. The first reduction efficiency varies by 0.5\% across the fiducial volume, the energy cut by 4\%, and the $lt$ cut by 5\%. The summary is listed in Table~\ref{tab:cmpspgl}. $T$ is the live time. 
Table~\ref{tab:cmpspgl} lists the measured $R_i$'s of radioactive isotopes over the entire energy spectrum. The $R_i$ of $^9$Li  was found to be strongly anti-correlated with both $^8$He and $^9$C due to their similar lifetimes.
This issue can be addressed by applying the neutron tagging method described in Section~\ref{sec:yieldntag}, since both $^9$C and $^8$He are unlikely to produce a neutron, especially if the energy is greater than 7.5 MeV~\cite{duke}. The relative fractions of $^8$B and $^8$Li could affect the fit;
however the $R_i$  of $^8$B and $^8$Li could not be decoupled due to their similar lifetimes and
endpoint energies ($\sim$13 MeV for $^8$Li and $\sim$14 MeV for $^8$B).
In the fit, the fractions of $^8$B and $^8$Li were assumed to be equal. By changing the relative fraction of $^8$Li from 0 to 1, the differences in the $R_i$'s were calculated and were included in the systematic uncertainty.

\begin{table}[thbp]
  \caption[Fit results of isotopes]{The measured production rates ($R_i$) over the entire energy spectrum for spallation-induced radioactive isotopes using the fit in the full time range. The first column lists the names of radioactive isotopes, the second column lists the
          selection efficiency $\epsilon_i$, and the third column lists the
          production rates of radioactive isotopes from the fit. The first and second uncertainties in $R_i$ represent the statistical and the systematic uncertainties, respectively. The systematic uncertainties include contributions from the data reduction and the maximum deviation of rates observed by changing the relative fraction of $^8$Li from 0 to 1. The fitted rates of $^8$He/$^9$C, $^{11}$Be and $^{15}$C are consistent with zero, so we set upper limits at 90\% confidence level (C.L.).
          }
  \label{tab:cmpspgl}
  \begin{tabular}{lccc} \hline
    \hline       Radioactive isotope        & $\epsilon_i$ & $R_i$ (kton$^{-1}$day$^{-1}$)\\
    \hline       ${}^{12}$B  &  45.5\%     &  19.8$\pm$0.1$\pm$1.0  \\
                 ${}^{12}$N  &  56.2\%       & 2.8$\pm$0.1$\pm$0.1  \\
                 ${}^{16}$N  &  45.0\%    & 39.7$\pm$3.3$\pm$2.8  \\
                 ${}^{11}$Be  & 38.1\%   &  $<$16.9 \\
                 ${}^{9}$Li   & 39.2\%     & 0.9$\pm$0.3$\pm$0.3  \\
                 ${}^{8}$He/${}^{9}$C    & 22.2\%, 50.2\% & $<$1.4  \\
                 ${}^{8}$Li/${}^{8}$B      &  42.8\%, 51.3\%      & 8.3$\pm$0.3$\pm$0.3\\
                 ${}^{15}$C                    &  31.8\%      & $<$6.7  \\\hline
    \hline
  \end{tabular}
\end{table}

\section{Improvement of the $^9$L\lowercase{i} yield measurement}
\label{sec:yieldntag}
The $^9$Li $\beta + n$ decay has the same signature as the
IBD reaction, so this irreducible cosmogenic background can only be statistically subtracted from the low energy IBD sample. Therefore, a precise measurement will greatly benefit searches for the IBD signal.
As the measurement of $^9$Li yield (see Section~\ref{sec:yielddt}) had large uncertainties, we improved it by tagging the neutrons from $^9$Li $\beta + n$ decays using the 2.2 MeV $\gamma$ ray emitted by the neutron capture on hydrogen.

\subsection{Neutron tagging}
Ref.~\cite{Zhang2015} introduced a neutron tagging method with a $\sim$∼20\% neutron detection efficiency and a $\sim$1\% probability that a background event would be misidentified as a neutron. 
This method incorporated eight separate cuts to distinguish the 2.2 MeV $\gamma$ rays from the accidental coincidence of PMT noise fluctuations. As the yield of $^9$Li
is low compared to other spallation backgrounds (which do not produce neutrons), this neutron capture selection was retuned and improved to get a larger rejection of accidental coincidence background.

\subsubsection{Discriminating variables for 2.2 MeV $\gamma$'s}
The PMT hits in the time range between 50 ns and 535 $\mu$s after the prompt event were used to search for the 2.2 MeV $\gamma$ rays from neutron captures. Each PMT hit time was corrected by the time of flight (TOF) of the photon from the
vertex of the prompt event to the PMT. (Using the prompt event vertex as an approximation for the neutron capture vertex is reasonable since the mean free path of a thermal neutron is similar to the vertex resolution at
10 MeV, and it is difficult to fully reconstruct low light yield events.) A 2.2 MeV $\gamma$ candidate was required to have more than 7 hits within 10 ns.

Besides the eight variables described in Ref.~\cite{Zhang2015}, 
five other variables were also developed as follows 
\begin{itemize}
\item [(I)] The mean value of the number of collected photo-electrons for the hit distribution;
\item [(II)] The root-mean-square (rms) of the number of collected photo-electrons for the hit distribution;
\item [(III)] The number of PMT hits inside a 20$^\circ$ cone around the sum vector of all PMT hit directions as seen from the vertex;
\item [(IV)] The number of PMT hits with collected charge above three photo-electrons, and
\item [(V)] The rms value of the angle between the individual and sum of hit direction vectors. 
\end{itemize}

\subsubsection{Optimization of neutron tagging}
\label{sec:ntag}
To optimize the neutron tagging, we adopted the multilayer perceptron
(MLP) method~\cite{TMVA2007, ROOT}. Unlike the previous analysis, the MLP method combined all the discriminating variables into one single output variable; the resulting accidental background level for the same acceptance was much lower than~\cite{Zhang2015}.
To train the MLP, we defined a background sample from a random trigger data regularly taken
for the purpose of evaluating the realtime noise level. A MC simulation of 2.2 MeV $\gamma$ rays with a neutron capture time
of 200 $\mu$s incorporating noise from the random trigger data served as the signal
sample.

We maximized the significance defined as $s/\sqrt{b}$ where $s(b)$ is the signal (background) level assuming an initial signal/noise ratio of 1. The optimal signal (background) efficiency was found to be 9.6\% (0.070\%). Both values refer to the time range from 50 ns to 535 $\mu$s.  
Compared with the previous neutron tagging method, this modified method suppressed accidental background by a factor of 16 at a cost of reducing the neutron tagging efficiency by only a factor of two. This signal efficiency $\epsilon_n$ was used in calculating the $^9$Li production rate. 

\subsubsection{Neutron Tagging Uniformity and Validation}
\label{sec:effn}
The time variation in the accidental background level was 23\% for the period of SK-IV and was mostly due to increasing PMT gain and noise rate. We assigned this variation as the systematic uncertainty. 

To test the neutron tagging uniformity, we subdivided the fiducial volume
into 10 bins in square of radius and 11 bins in height (for definitions, see Ref.~\cite{Abe2011}).
The variation in signal (background) efficiency was found to be 20\% (30\%).
The time distribution of events selected by the neutron tagging from the random trigger data was observed to be flat, and differs from the expected exponential distribution from neutron captures in water. 

To validate the neutron tagging efficiency as well as its position dependence, an
Am/Be source was deployed at three different positions with Source A at the
center of the ID, Source B close to the barrel of the ID, and Source C close to the top of the ID.
More description of the experimental setup and other details can be found 
elsewhere \cite{Zhang2015, ntag2009}.
Table~\ref{tab:source_position} lists the detailed positions of the sources, together 
with a comparison between the neutron tagging efficiency $\epsilon_n$(Am/Be) 
obtained from the Am/Be neutron source and that obtained from the MC simulation.
\begin{table}[thbp]
  \caption[Am/Be source position]{The positions of Am/Be for Source A, B, C. The last two
  columns give a comparison between the neutron
  tagging efficiency $\epsilon_n$(Am/Be) obtained from the Am/Be neutron source
  and that obtained from the MC simulation. See text for more details.}
  \label{tab:source_position}
  \begin{center}
  \begin{tabular}{cccccc}
    \hline \hline   Source   & X (cm) & Y (cm) & Z (cm) & $\epsilon_n$(Am/Be)(\%) &$\epsilon_n$(MC)(\%)\\
    \hline
      A                   & 35.3 & -70.7 & 0.      &$10.1\pm0.2$ & $ 10.2$\\

      B                   & 35.3 & -1201.9 & 0.  &$12.0\pm0.4$ & $10.2$ \\

      C                   & 35.3 & -70.7 & 1500.&$11.3\pm0.4$ & $11.3$ \\
    \hline \hline
  \end{tabular}
   \end{center}
\end{table}

The combined neutron capture time in water was measured to be ($202.6\pm3.7$) $\mu$s, which was in good
agreement with the (204.7$\pm$0.4) $\mu$s in Ref.~\cite{ncapturelife}. 
Combining the neutron tagging efficiencies of the three source positions gave a weighted average of $(10.6\pm0.2$)\%, which was compared to the neutron tagging efficiency
$\epsilon_n$ in Section~\ref{sec:ntag}.The maximum percent difference between the optimal signal efficiency found in Section~\ref{sec:ntag} (9.6\%) and the neutron tagging efficiency determined using the Am/Be source was calculated to be 10.4\%. This value was used as the overall systematic uncertainty for the neutron tagging efficiency.

\subsection{$^9$Li rate and yield}
After validating the modified neutron capture selection, we selected $^9$Li decay candidates with a $\beta$$+$$n$ pair from the spallation sample with a 7.5 MeV trigger threshold for the low-energy events associated with the preceding muons:
\begin{itemize}
\item [(I)] Triplets ($\mu$, $\beta$, $n$) must pass the first reduction described in Section~\ref{sec:yielddt}.
\item [(II)] Events should satisfy 0.05 s $<dt<$ 0.5 s and $lt<$  200 cm to reject accidental coincidences and longer-lived spallation backgrounds. 
\item [(III)] No additional muon within 1 ms was allowed to remove most of multiple neutron events induced by the preceding muons.
\item [(IV)] Detected energy was required to be between 7.5 and 14.5 MeV. 
\end{itemize}

\subsubsection{The $\beta$ energy}
\label{sec:bkgli9}
\begin{figure}[!t]
   \centering
  \includegraphics[width=0.9\columnwidth]{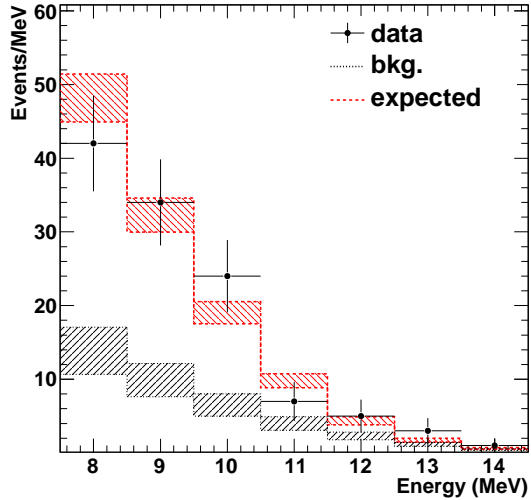}
  \renewcommand{\figurename}{Fig.}
   \caption{Detected $\beta$ energy spectrum (points) for all the $^9$Li $\beta$$+$$n$ decay candidates. Also shown are the expected spectrum (shaded areas with dashed lines) using the electron spectrum from $^9$Li $\beta$$+$$n$ decay modes and the spectrum of accidental background.}
   \label{fig:spenli}
 \end{figure}

The lower bound of the detected energy range was limited by
the trigger threshold. The live time for this sample is 998 days.
In total, we observed 116 candidates for the decay of $^9$Li.
No event with multiple neutrons was observed in these candidates. Among these candidates, we estimated the number of events without 
neutrons to be 38.2 with the total 54,963 $\beta$ events scaled by the accidental background level of 0.070\% introduced by the neutron tagging (see Section~\ref{sec:ntag}). This approximative calculation neglected the small component 
from the $^9$Li $\beta$$+$$n$ decays in the $\beta$ event sample. 
We also treated the spectrum for the accidental background in the 
same manner. 
Fig.~\ref{fig:spenli} shows the detected $\beta$ energy spectrum for all the 
$^9$Li $\beta$$+$$n$ decay candidates. Also shown are the expected spectrum from the 
sum of both the signal and accidental background spectra. 
The observed and the expected spectra were consistent within the uncertainties.  
We found a clear event excess compared with 
the background-only scenario (points over the shaded areas with dotted line). The number of $^9$Li $\beta$$+$$n$ decay
events was estimated to be $(77.8\pm10.8)$ events, which was obtained by subtracting the estimated number of events without neutrons from the number of $^9$Li candidates. The systematic uncertainty due to the accidental background estimate was evaluated to be 8.8 events, which gave a relative difference of 11.2\%. 

\subsubsection{The neutron capture time} 
\begin{figure}[!t]
   \centering
  \includegraphics[width=0.9\columnwidth]{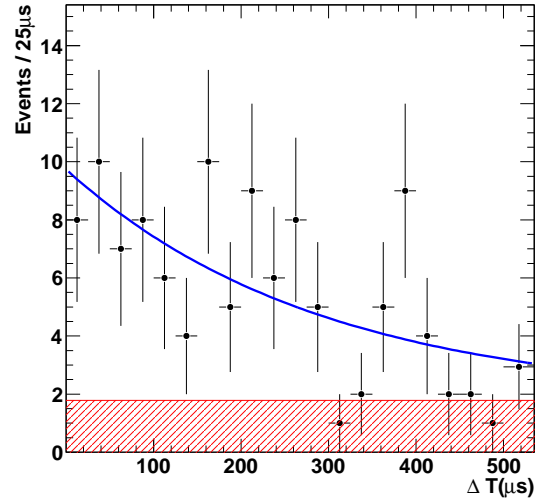}
  \renewcommand{\figurename}{Fig.}
   \caption{The distribution of time differences between the $\beta$ and the 2.2 MeV $\gamma$ for the 
$^9$Li $\beta$$+$$n$ decay candidates (points). The solid line shows the fit result, while 
the shaded region represents the 
expected backgrounds without neutrons.}
   \label{fig:spdtli}
 \end{figure}

As a cross check to the background estimate given above, we studied the neutron capture time
from the time difference between the $\beta$ and the 2.2 MeV $\gamma$ for the 
$^9$Li $\beta$$+$$n$ decay candidates. 
A fit with a fixed neutron capture time at 200 $\mu$s and a free background level 
was performed to the distribution of time differences, giving a fitted 
background of $55.6\pm16.7$ events which agreed with the 
estimate given in Section~\ref{sec:bkgli9} within one standard deviation. 
As shown in Fig.~\ref{fig:spdtli}, we also conducted an unbinned maximum likelihood fit
with an exponential plus a constant background estimate (see Section~\ref{sec:bkgli9}). The fitted capture time was ($281.2\pm77.5$) $\mu$s,
which was consistent with the expectation, indicating that events 
with $\beta$$+$$n$ pairs were observed. 

\subsubsection{The $^9$Li lifetime}
\begin{figure}[!t]
   \centering
  \includegraphics[width=0.9\columnwidth]{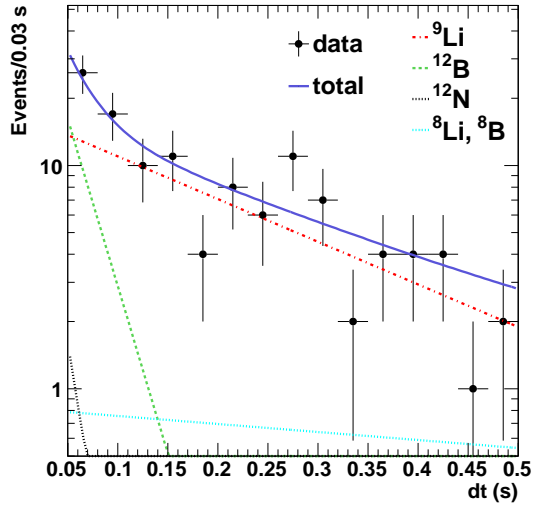}
  \renewcommand{\figurename}{Fig.}
   \caption{The distributions of the time difference 
   between the preceding muon and the $\beta$ for the
$^9$Li $\beta$$+$$n$ decay candidates.
The points represent the data. The lines correspond to various radioactive isotopes and their total. 
The contributions from $^{15}$C, $^{16}$N, $^{11}$Be and $^{8}$He/$^{9}$C are not shown, since 
they are tiny as indicated in Table~\ref{tab:estimatebs}. 
   }
   \label{fig:dtlia}
 \end{figure}

Another cross check was the $^9$Li lifetime, which was defined by the time difference between 
the preceding muon and the $\beta$ for the
$^9$Li $\beta$$+$$n$ decay candidates.  Fig.~\ref{fig:dtlia} shows 
the $^9$Li lifetime distribution. All the accidental 
background sources were understood to be from $^{12}$B, $^{12}$N, $^{15}$C, $^{8}$Li/$^{8}$B, $^{16}$N, $^{11}$Be and $^{8}$He/$^{9}$C decays together with the $^{9}$Li $\beta$ decay without a neutron.  We calculated the number of events for each background source based on $R_i$ from 
Table~\ref{tab:cmpspgl} using
\begin{equation}
\label{eq:nbes}
N_b = R_i \cdot FV \cdot  T \cdot \epsilon_b, 
\end{equation}
where $T$ is the live time of the data. 
$\epsilon_b$ is the remaining background level after the first reduction, $dt$ cut, $lt$ 
cut, energy cut and neutron tagging cut. 
The quantities of the variables in Eq.~(\ref{eq:nbes}) can be found in Table~\ref{tab:estimatebs}. 
\begin{table}[thbp]
  \caption{The quantities of the variables in Eq.~(\ref{eq:nbes}).}
  \label{tab:estimatebs}
  \centering
  \begin{tabular}{lccccccc} \hline
    \hline    Radioactive isotope   & $\epsilon_b$               & $N_b$   &   \\ \hline
              $^{9}$Li w/o $n$          &  1.1$\times10^{-4}$     &     2.3  &  \\
              $^{12}$B                      &   3.5$\times10^{-5}$    &     15.6  &  \\
              $^{12}$N                      &   1.4$\times10^{-5}$    &     0.9  &  \\
              $^{15}$C                      &   7.8$\times10^{-6}$    &     0.6  &  \\
              $^8$Li/$^{8}$B            &   5.3$\times10^{-5}$     &     9.9  &  \\
              $^{16}$N                     &   5.2$\times10^{-6}$     &     4.6   &  \\ 
              $^{11}$Be                   &   3.1$\times10^{-6}$      &    0.4   &  \\ 
              $^{8}$He/$^{9}$C       &  3.9$\times10^{-5}$       &    0.4   &  \\ \hline
              Total                            &    -                                 &     34.7  &  \\
    \hline \hline
  \end{tabular}
\end{table}
The total number of 34.7 expected events from the eight background sources given in Table~\ref{tab:estimatebs} was 
in good agreement with the estimate given in Section~\ref{sec:bkgli9}. 
Fixing the lifetimes of background isotopes listed in Table~\ref{tab:isotope} and their expected events 
given in Table~\ref{tab:estimatebs}, we performed a one-parameter fit to the lifetime distribution ($dt$) 
with eight exponentials. The fitted lifetime of $^9$Li was ($0.23\pm0.06$) s, which 
was also in good agreement with the expected 0.26 s.
This indicated that the $^9$Li events were observed 
via the neutron tagging method at SK-IV.

We noted that the cosmic-ray muon induced $^8$He also has a $\beta$$+$$n$ decay mode  
with a branching ratio of $\sim$16\%~\cite{Tilley}. However, the fraction of $\beta$ from 
the $^8$He $\beta$$+$$n$ decays above the trigger threshold
was $<$1\%, and thus was ignored. The contribution of muon-induced $^9$C was also ignored because there is no neutron produced in this decay.

\subsubsection{$^9$Li production rate}
After performing cross checks using both the neutron capture time and $^9$Li lifetime distributions,   
we calculated the rate using
\begin{equation}
R_{{^9}\text{Li}} = \frac { N_{{^9}\text{Li}} } {  FV\cdot BR\cdot T\cdot \epsilon_{^9Li} },
\label{eq:nn}
\end{equation}
where N$_{{^9}\text{Li}}$ is the number of $^9$Li decay events in Section~\ref{sec:bkgli9}, $BR=50.8$\% is the branching ratio for the $^9$Li $\beta$$+$$n$ decay mode~\cite{Tilley}. 
$\epsilon_{^9Li}$ is the efficiency, including 16.3\% for the fraction of electrons above the trigger threshold from the $^9$Li $\beta$$+$$n$ decays; 
4.9\% for the first reduction, $dt$ and $lt$ cuts, and neutron tagging cuts. 
The  $^9$Li production rate was finally measured to be
\begin{equation}
\label{eq:Rli9}
 R_{{^9}\text{Li}} =(0.86\pm0.12_{\text{stat.}})~ \text{kton}^{-1}\text{day}^{-1}. 
\end{equation} 

\subsubsection{Systematic uncertainties}
The systematic uncertainties are summarized in Table~\ref{tab:systli9}.
The uncertainties on the live time calculation and the efficiency of first reduction were taken from Ref.~\cite{Bays2012}. 
Since the precision of the time measurements is much better than the uncertainty of the $^9$Li lifetime, the uncertainty on 
the $dt$ cut was ignored. As for the $lt$ cut, the efficiency varies for different muon classes due to their 
different $lt$ distributions as already shown in Fig.~\ref{fig:dtdl}. 
In particular, the $lt$ distribution of multiple muons is much wider than the others.
According to the theory~\cite{li2015}, all the spallation isotopes including $^9$Li were made in muon showers. 
In addition to the systematic study in Section~\ref{subsec:disdt}, we studied the production rate specifically for the multiple 
muon events. We  observed that about 40\% of $^9$Li's are indeed produced by these events. The systematic uncertainty related to the $lt$ cut is estimated using data, as discussed in Section~\ref{subsec:disdt}.
The systematic uncertainties for the neutron tagging efficiency and the accidental background level were estimated in Section~\ref{sec:effn}
and Section~\ref{sec:bkgli9}. Since we had an energy threshold cut on the $^9$Li energy tail, the difference between the true and the 
simulated energy spectra could vary with the water transparency, PMT gain and dark noise, etc.  These effects 
were taken into account in MC simulation; the maximum relative deviation in the efficiency for the energy threshold 
was found to be 8.0\% and assigned as the systematic uncertainty.  
The uncertainty of $BR$ was obtained from Ref.~\cite{Tilley}.
The total systematic uncertainty was added in quadrature and was found
to be 18.1\%. 
\begin{table}[thbp]
 \caption[The breakdown of systematic uncertainties for the ${}^{9}$Li
 measurement with neutron tagging]{The break down of relative systematic uncertainties for the ${}^{9}$Li
 measurement with the neutron tagging.}
 \label{tab:systli9}
 \centering
 \begin{tabular}{lccc} \hline
   \hline    Source   &   Systematic uncertainty    \\
   \hline 
             Live time        & 0.1\%  \\  
             First reduction  & 1.0\%  \\
            $lt$ reduction  & 4.8\%  \\
             Energy threshold   & 8.0\%  \\                 
             n-tag efficiency   & 10.4\% \\
             Bkg. estimate        & 11.2\%  \\
            $BR$  & 1.8\%  \\ \hline
            Total  & 18.1\%  \\\hline \hline
 \end{tabular}
\end{table}

\begin{table*}[thbp]
 \caption{Measured spallation-induced radioactive isotope yields ($Y_i$'s) in SK.  The fitted yields of $^8$He/$^9$C, $^{11}$Be and $^{15}$C are consistent with zero, so we set upper limits at 90\% C.L..
 The yields from a theoretical calculation  \protect{\cite{li2014}}  are shown in the third column.  KamLAND's experimental and simulation results ~\cite{kamland2010}  are listed in the fourth and fifth columns, respectively. 
  The unit is  $10^{-7}\mu^{-1}g^{-1}cm^2$.}
  \label{tab:metheo}
  \centering
  \begin{tabular}{lcccc} \hline
    \hline       Isotope        & $Y_i$ in water & Expected~\cite{li2014}  &  ~~$Y_i$ in scintillator~\cite{kamland2010}   & ~~Expected~\cite{kamland2010} \\\hline
                 ${}^{12}$B &  11.7$\pm$0.1$\pm$0.6       &  12        &  42.9$\pm$3.3  & 27.8$\pm$1.9 \\
                 ${}^{12}$N   & 1.6$\pm$0.1$\pm$0.1      &  1.3         & 1.8$\pm$0.4    &  0.77$\pm$0.08\\
                 ${}^{16}$N  & 23.4$\pm$1.9$\pm$1.7     &  18          &  -                      &   -  \\
                 ${}^{11}$Be   & $<$10.0      &  0.81      &  1.1$\pm$0.2   & 0.84$\pm$0.09\\
                 ${}^{9}$Li w/o n-tag   & 0.5$\pm$0.2$\pm$0.2     &   1.9   &  &  \\
                 ${}^{9}$Li w/ n-tag    &   0.51$\pm$0.07$\pm$0.09    &   1.9   &  2.2$\pm$0.2  &  3.16$\pm$0.25 \\
                 ${}^{8}$He/${}^{9}$C   &  $<$0.9     & 1.1  & 0.7$\pm$0.4/3.0$\pm$1.2 & 0.32$\pm$0.05/1.35$\pm$0.12\\
                 ${}^{8}$Li/${}^{8}$B,   & 4.9$\pm$0.2$\pm$0.2       & 18.8 &   12.2$\pm$2.6/8.4$\pm$2.4 & 21.1$\pm$1.4/5.7$\pm$0.4   \\
                 ${}^{15}$C   & $<$3.9      & 0.82    &  -  &  -  \\\hline
    \hline
  \end{tabular}
\end{table*}

\section{RESULTS AND DISCUSSIONS}
With the quantities listed in Section~\ref{sec:muons}, Section~\ref{subsubsec:produc}, and Table~\ref{tab:cmpspgl}, 
the yield of the $i$-th radioactive isotope ($Y_i$) is calculated by
\begin{equation}
\label{eq:yiall}
Y_i= \frac{R_i \cdot FV }{R_{\mu} \cdot \rho \cdot L_\mu},
\end{equation}
where $\rho$ is the density of water.
The $^9$Li yield measurement using the neutron tagging is calculated from 
the $^9$Li production rate given in Eq.~(\ref{eq:Rli9}) and the
corresponding systematic uncertainties are 
listed in Table~\ref{tab:systli9}. 
The measured results and a theoretical calculation are given in Table~\ref{tab:metheo}. 
The theoretical calculation is given by Ref.~\cite{li2014}, 
which is based on the simulation code FLUKA~\cite{Ferrari, Battistoni}. 
\label{sec:discuss}

This is the first measurement of the radioactive isotope yields
produced by cosmic-ray muons in underground water detectors.  
The mean values of the $^{12}$B, $^{12}$N and $^{16}$N theoretical yield calculations are inside the two-sigma confidence interval of the respective measurements. The best-fit $^{12}$B yield is only 2.5\% smaller than the calculation while the corresponding $^{8}$Li yield measurement ~\cite{kamland2010} (see Table~\ref{tab:metheo}) in scintillator is 42\% below expectation. The best-fit $^{12}$N yield is 23\% larger than the calculation while the corresponding $^{8}$B yield measurement in scintillator
exceeds the calculation by 47\%. The best-fit $^{16}$N yield is 30\% larger than the calculation, the corresponding $^{12}$B yield measurement in scintillator
exceeds the calculation by 54\%. The upper limit of the $^8$He/$^9$C yield is only 10\% lower than the theoretical prediction while the upper limits of $^{11}$Be and $^{15}$C are consistent. However, we found that the yields of the spallation ejecting many nucleons  of the $^{16}$O into $^8$Li/$^8$B and $^9$Li are a factor of 3.6$-$4.1 and 3.1$-$4.7 below the calculations. There is no corresponding spallation of the $^{12}$C nucleus in scintillator, since $^4$Li is very short-lived, and $^4$H and $^5$H do not exist. The larger discrepancy could be owed to the spallation ejecting many nucleons, or a feature of the $\pi$-initiated reaction. While scintillator measurements show agreement for those isotopes within a factor of two, isotopes produced by spallation ejecting many nucleons of $^{12}$C are not measured. We also note that the two different measurements on the $^9$Li yield
in SK are in good agreement, and the one with neutron tagging is better
because of the strong background suppression.
  
Since the production rates of the radioactive isotopes depend on a power law
of the muon energy~\cite{kamland2010}, the measured yields, especially the $^{9}$Li yield,
can be directly applied to the evaluation of the spallation background
in SK and future experiments, such as the Super-Kamiokande gadolinium project
(SuperK-Gd)~\footnote{Ref.~\cite{gdzooks!} proposed adding a 0.2\% gadolinium 
solution into the SK water. After exhaustive studies, on June 27, 2015, the SK 
Collaboration formally approved the concept, officially initiating the SuperK-Gd project, which will enhance anti-neutrino detectability (along with other physics capabilities) by dissolving 0.2\% gadolinium sulfate by mass in the SK water.}, Hyper-Kamiokande~\cite{hk}, etc.

In SuperK-Gd, the neutron tagging efficiency will be increased 
to about 80\% (90\% capture and 90\% reconstruction efficiencies) 
when 0.2\% of Gd$_2$(SO$_4$)$_3$ by mass is added into the SK water.
Assuming 80\% IBD detection efficiency, we expect 0.5$-$8.1 SRN signal events /22.5 kton/year 
in 9.5$-$29.5 MeV positron kinetic energy range for various theoretical models~\cite{Ando03, Totani95, Malaney97, Hartmann97, Kaplinghat00, Strigari03, Totani96, HBD09, Fukugita03, Lunardini09}. 
From our measurement on the $^9$Li yield, assuming 
80\% neutron tagging efficiency and 
0.5\% probability of the muon-induced $^9$Li event leakage after the
spallation cuts, the expected background events from the cosmogenic
$^9$Li would become $0.5\pm0.1\pm0.2$  events/22.5 kton/year in the same energy window. 
 
Furthermore, as the spectral shapes of the $^9$Li  background and the SRN signal differ -- most 
of the $^9$Li events are found in the bottom three 1~MeV bins, while the predicted SRN spectra 
are much flatter -- the signal to noise ratio can be significantly optimized by adjusting the 
lower edge of the allowed energy range upward slightly.  Such an energy cut will also serve to shield against  occasional highly upward-fluctuated reactor anti-neutrinos.  The study presented herein  
therefore demonstrates for the first time that spallation-induced $^9$Li will have a negligible effect on 
SuperK-Gd's SRN search. 



\section{SUMMARY}
\label{sec:summary}
In summary, the first measurement of the yields of radioactive isotopes produced by
cosmic-ray muons in an underground water Cherenkov detector 
was performed using time distributions in SK-IV data. 
The path-length distribution of cosmic-ray muons, 
the relative time and distance distributions between parent muons and
daughter isotopes,
the energy distribution of $^9$Li $\beta$$+$$n$ decays, 
and so on are expected to facilitate further insights into the mechanism of spallation in water, 
which should be useful in developing a more practically powerful tool to reject
cosmogenic backgrounds in solar, reactor and supernova relic neutrino
experiments.
Finally, the yield of $^9$Li was more precisely measured by tagging neutrons.
This information regarding an important potential background source could aid 
in the design and optimization of future water Cherenkov detectors that aim to search for SRNs. Agreement of the data with theoretical calculations is in general much better in water than in scintillator; only those isotopes produced as a result of ejecting many nucleons of $^{16}$O significantly deviate - by about a factor of four - from predictions.

\section{ACKNOWLEDGMENT}
We gratefully acknowledge the cooperation of the Kamioka Mining and Smelting Company. The Super-Kamiokande experiment has been built and operated from funding by the Japanese Ministry of Education, Culture, Sports, Science and Technology, the U.S. Department of Energy, and the U.S. National Science Foundation. Some of us have been supported by funds from the Research Foundation of Korea (BK21 and KNRC), the Korean Ministry of Science and Technology, the National Research Foundation of Korea (NRF- 20110024009 ), 
the European Union FP7 (DS laguna-lbno PN-284518 and ITN invisibles GA-2011-289442), the Japan Society for the Promotion of
Science, the National Science and Engineering Research Council (NSERC) of Canada, and the Scinet and Westgrid consortia of Compute Canada; This work is also supported by the National Natural Science Foundation of China (Grants No.11235006).



\begin{thebibliography}{}
\bibitem{hagner2000}
T.~Hagner {\it et al.}, Astropart. Phys. 14, 33(2000).
\bibitem{kamland2010}
S.~Abe {\it et al.}, (KamLAND Collaboration), Phys.Rev. C81, (2010) 025807.
\bibitem{bxno2013}
B. Bellini {\it et al.}, (Borexino Collaboration), J. Cosmol. Astropart. Phys. 08 (2013) 049.
\bibitem{gdzooks!}
J. F. Beacom and M. R. Vagins, Phys. Rev. Lett., {\bf 93}: 171101 (2004). 
\bibitem{hk}
K.~Abe, {\it et al.}, (Hyper-Kamiokande working group), arXiv:1109.3262.
\bibitem{watchman}
M. Askins, {\it et al.}, (WATCHMAN Collaboration), arXiv:1502.01132.
\bibitem{memphys}
L. Agostino, {\it et al.}, (MEMPHYS Collaboration), J. Cosmol. Astropart. Phys. 1301 (2013) 024.
\bibitem{lbne}
C. Adams, {\it et al.}, (LBNE Collaboration), arXiv:1307.7335. 
\bibitem{Fukuda2001}
S.~Fukuda {\it et al.}, (The Super-Kamiokande Collaboration), Phys. Rev. Lett. 86, 5651 (2001).
\bibitem{Hosaka2006}
J.~Hosaka {\it et al}, (Super-Kamiokande Collaboration), Phys. Rev. D 73, 112001 (2006).
\bibitem{Abe2011}
K.~Abe {\it et al.}, (The Super-Kamiokande Collaboration), Phys. Rev. D 83, 052010 (2011).
\bibitem{Bays2012}
K.~Bays {\it et al.}, (The Super-Kamiokande Collaboration), Phys. Rev. D 85, 052007 (2012).
\bibitem{li2014}
S. W. Li and J. F. Beacom, Phys. Rev. C 89, 045801 (2014).
\bibitem{li2015}
S. W. Li and J. F. Beacom, Phys. Rev. D 91, 105005 (2015).
\bibitem{li20152}
S. W. Li and J. F. Beacom, arXiv:1508.05389.
\bibitem{Tilley}
D. R.~Tilley, J. H. Kelley, J. L. Godwin, D. J. Millener, J. E. Purcell, C. G. Sheu, and H. R. Weller, Nucl. Phys. A 745, 155 (2004).
\bibitem{Zhang2015}
H.~Zhang {\it et al.}, (Super-Kamiokande Collaboration), Astropart. Phys. 60 (2015) 41.
\bibitem{SK}
K.~Abe, {\it et al.}, (Super-Kamiokander Collaboration), Nucl. Instr. Meth, A 737 (2014).
\bibitem{SK4ONLINE}
S. Yamada, {\it et al.},  IEEE Trans. Nucl. Sci.,  57 (2010) 428.
\bibitem{muon1}
Z.~Conner, PhD thesis, University of Maryland, 1997.
\bibitem{muon2}
S.~Desai, PhD thesis, Boston Univesity, 2004.
\bibitem{nndc}
http://www.nndc.bnl.gov.
\bibitem{duke}
http://www.tunl.duke.edu/nucldata/GroundStatedecays/.
\bibitem{Geant3}
R. Brun, R. Hagelberg, M. Hansroul, and J. C. Lassalle, Simulation program for particle physics experiments, GEANT : user guide and reference manual, CERN-DD-78-2(1978).
\bibitem{TMVA2007}
        A.~Hoecker, P.~Speckmayer, J.~Stelzer,
        J.~Therhaag, E.~von Toerne, and H.~Voss,
        PoS A CAT 040 (2007); arXiv: physics/0703039.
\bibitem{ROOT}
P.~Speckmayer, A.~Hoecker, J.~Stelzer, and H.~VossJ. Phys.: Conf. Ser. 219, 032057 (2010).
\bibitem{ntag2009}
H. Watanabe {\it et al.}, (Super-Kamiokande Collaboration), Astropart. Phys., 2009, {\bf 31}: 320.
\bibitem{ncapturelife}
D. Cokinos and E. Melkonian, Phys. Rev. C 15, 1636 (1977). 
\bibitem{Ferrari}
A. Ferrari, P. R. Sala, A. Fasso, and J. Ranft, FLUKA: A multi-particle transport code, CERN-2005-10 (2005),
INFN-TC-05-11, SLAC-R-773.
\bibitem{Battistoni}
G. Battistoni {\it et al.}, AIP Conf. Proc. 896, 31 (2007).
\bibitem{Ando03}
S.~Ando, K.~Sato and T. Totani, Astropart. Phys. 18, 307 (2003).
\bibitem{Totani95}
T. Totani and K. Sato, Astropart. Phys. 3, 367 (1995).
\bibitem{Malaney97}
R. A. Malaney, Astropart. Phys. 7, 125 (1997).
\bibitem{Hartmann97}
D. H. Hartmann and S. E. Woosley, Astropart. Phys. 7, 137 (1997).
\bibitem{Kaplinghat00}
M. Kaplinghat, G. Steigman, and T. P. Walker, Phys. Rev. D 62, 043001 (2000).
\bibitem{Strigari03}
L. Strigari, M. Kaplinghat, G. Steigman, and T. Walker, JCAP, 0403:007 (2004).
\bibitem{Totani96}
T. Totani, K. Sato, and Y. Yoshii, Astrophys. J. 460, 303 (1996).
\bibitem{HBD09}
S. Horiuchi, J. F. Beacom, and E. Dwek, Phys. Rev. D 79, 083013 (2009).
\bibitem{Fukugita03}
M. Fukugita and M.Kawasaki, Mon. Not. Roy. Astron. Soc. 340, L7 (2003).
\bibitem{Lunardini09}
C. Lunardini, Phys. Rev. Lett. 102, 231101(2009).
\end{thebibliography}
\end{document}